\documentclass[aip,a4paper,prl,twocolumn,reprint,superscriptaddress]{revtex4-1}
\usepackage{graphicx}
\usepackage[english]{babel}
\usepackage{amsmath}
\usepackage{amsfonts}
\usepackage{amssymb}
\usepackage{color}
\usepackage[normalem]{ulem}

\usepackage{multibib}           
\newcites{Me}{Methods references}            

\date{\today}

\begin{document}

\renewcommand{\figurename}{\textbf{Fig.}}

\title{\Large Spanning Fermi arcs in a two-dimensional magnet}
\author {Ying-Jiun Chen}
\altaffiliation{Corresponding author, E-mail: yi.chen@fz-juelich.de}
\affiliation {Peter Gr\"unberg Institut, Forschungszentrum J\"ulich, 52425 J\"ulich, Germany}
\affiliation {Fakult\"at f\"ur Physik, Universit\"at Duisburg-Essen, 47057 Duisburg, Germany}

\author {Jan-Philipp Hanke}
\affiliation {Peter Gr\"unberg Institut, Forschungszentrum J\"ulich, 52425 J\"ulich, Germany}
\affiliation {Institute for Advanced Simulation, Forschungszentrum J\"ulich and JARA, 52425 J\"ulich, Germany}

\author {Markus Hoffmann}
\affiliation {Peter Gr\"unberg Institut, Forschungszentrum J\"ulich, 52425 J\"ulich, Germany}
\affiliation {Institute for Advanced Simulation, Forschungszentrum J\"ulich and JARA, 52425 J\"ulich, Germany}

\author {Gustav Bihlmayer}
\affiliation {Peter Gr\"unberg Institut, Forschungszentrum J\"ulich, 52425 J\"ulich, Germany}
\affiliation {Institute for Advanced Simulation, Forschungszentrum J\"ulich and JARA, 52425 J\"ulich, Germany}

\author {Yuriy Mokrousov}
\affiliation {Peter Gr\"unberg Institut, Forschungszentrum J\"ulich, 52425 J\"ulich, Germany}
\affiliation {Institute for Advanced Simulation, Forschungszentrum J\"ulich and JARA, 52425 J\"ulich, Germany}

\affiliation {Institute of Physics, Johannes Gutenberg University Mainz, 55099 Mainz, Germany}

\author {Stefan Bl\"ugel}
\affiliation {Peter Gr\"unberg Institut, Forschungszentrum J\"ulich, 52425 J\"ulich, Germany}
\affiliation {Institute for Advanced Simulation, Forschungszentrum J\"ulich and JARA, 52425 J\"ulich, Germany}

\author {Claus M. Schneider}
\affiliation {Peter Gr\"unberg Institut, Forschungszentrum J\"ulich, 52425 J\"ulich, Germany}
\affiliation {Fakult\"at f\"ur Physik, Universit\"at Duisburg-Essen, 47057 Duisburg, Germany}
\affiliation {Department of Physics, University of California Davis, Davis, CA 95616, USA}

\author {Christian Tusche}
\altaffiliation{E-mail: c.tusche@fz-juelich.de}
\affiliation {Peter Gr\"unberg Institut, Forschungszentrum J\"ulich, 52425 J\"ulich, Germany}
\affiliation {Fakult\"at f\"ur Physik, Universit\"at Duisburg-Essen, 47057 Duisburg, Germany}

\begin{abstract}

The discovery of topological states of matter has led to a revolution in materials
research. When external or intrinsic parameters break certain symmetries, global properties of topological materials change drastically. A paramount example is the
emergence of Weyl nodes under broken inversion symmetry, acting like magnetic
monopoles in momentum space. However, while a rich variety of non-trivial quantum phases could in principle also originate from broken time-reversal symmetry, realizing systems that combine magnetism with complex topological properties is remarkably elusive due to both considerable experimental and theoretical challenges.
Here, we demonstrate that giant open Fermi arcs are created at the surface of ultrathin hybrid magnets. The Fermi-surface topology of an atomically thin ferromagnet is substantially modified by
the hybridization with a heavy-metal substrate, giving rise to Fermi-surface discontinuities that are bridged by the Fermi arcs. Due to the interplay between magnetism and topology, we can control both the shape and the location of the Fermi arcs by tuning the magnetization direction. The hybridization points in the Fermi surface can be attributed to a non-trivial ``mixed" topology and induce hot spots in the Berry curvature, dominating spin and charge transport as well as magneto-electric coupling effects. 

\end{abstract}

\maketitle

Classifying topological states of matter based on their global properties has matured into a pervasive scientific concept that provides an elegant interpretation of physical phenomena. Prominent examples include the recent discovery of axions in condensed matter~\cite{Li2019,Zhang2019}, the quantum anomalous Hall effect~\cite{yu2010quantized,chang2013experimental}, and the emergence of robust spin textures~\cite{Muehlbauer2009,hoffmann2017antiskyrmions}, 
all of which root in non-trivial characteristics of the wave function in magnetic solids. 
Ever since these striking advances, the intricate interplay between magnetism and topology is moving steadily into the focus of research, owing to bright promises for energy-efficient information processing~\cite{Fert2013,Han17} and brain-inspired computing~\cite{Zazvorka2019}. To achieve these innovative functionalities, realizing non-trivial topological phases in two-dimensional (2D) magnets opens up a highly attractive route for, e.g., the low-dissipation control of magnetism mediated by ``mixed" Weyl points~\cite{Han17}. Among such classes of intriguing 2D quantum materials are in particular layered van der Waals ferromagnets~\cite{Deng2018,Kim2018} and magnetic topological insulators~\cite{Fan2014,Niu19}, which received considerable attention lately.

Symmetries play a key role in identifying and understanding emergent phases of matter with complex topologies. For instance, time-reversal (TR) or inversion symmetries can manifest in protected exotic states in the electronic structure of topological insulators and Dirac semimetals~\cite{tan19,ver19,zha19}. If one of these protective symmetries is broken, the global properties of the material may change drastically as reflected by, for example, the splitting of individual Dirac points into pairs of Weyl points with opposite chirality. These points are linked by open Fermi arcs that emerge at the surface as a direct consequence of the non-trivial Fermi-surface topology~\cite{Xu613,Hal14,min19,bra16,rao19}. Since ferromagnetism spontaneously breaks TR symmetry, controlling the magnetic order serves as an ideal means to create on-demand topological phase transitions and significantly alter the topology of the electronic states.
Consequently, the concept of magnetic topological semimetals has recently attracted great research	interest~\cite{Mor19,Niu19}. While the systems studied so far still belong to the class of three-dimensional (3D) Weyl semimetals with topological points located in 3D momentum space, the concept needs to be extended for low-dimensional systems.
Despite the increasing importance of 2D magnetic materials~\cite{Deng2018,Kim2018,bur18,che19,mak19,Niu19}, very little is known about the interplay between magnetism and topology in such low-dimensional magnets.

%
%
\begin{figure*}[htb]
	\centering{\includegraphics[clip,width=16.6cm]{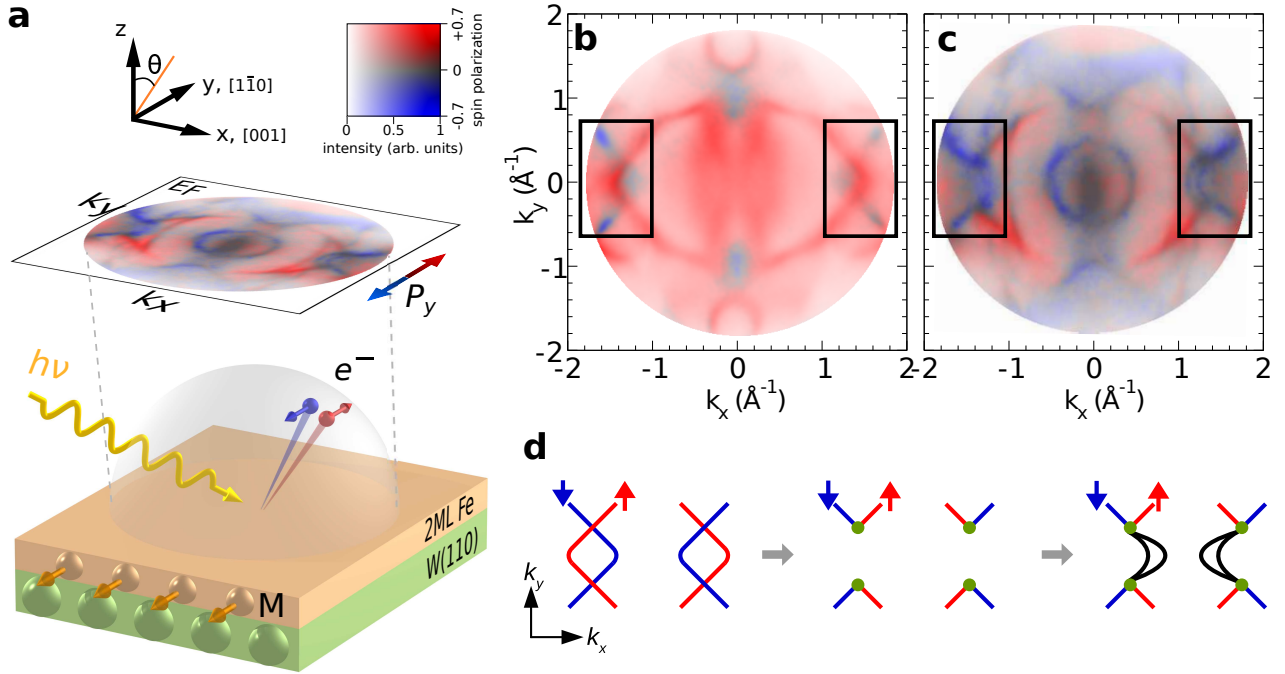}}
	\caption{\textbf {Fermi arcs in a 2D topological ferromagnet.}
		\textbf{a}, Experimental geometry for the spin-, orbital- and momentum-resolved photoelectron study of the Fermi surface. The magnetization {$\mathbf M$} and the spin polarization P$_{y}$ are oriented in-plane along the $y$ axis, i.e., the crystallographic $[1\bar 1 0]$ direction. \textbf{b}, \textbf{c}, Spin-resolved photoemission momentum maps at the Fermi energy $\textit{E}_{F}$ for (\textbf{b}) 12 MLs and (\textbf{c}) 2 MLs Fe films grown on W(110), measured with a photon energy of $h\nu$=50\,eV for a sample magnetization pointing into $-y$ direction. The spin polarization $P_y$ is indicated by red and blue colours, and the colour strength encodes the intensity. \textbf{d}, Schematic evolution of the Fermi-surface topology for the appearance of surface arcs upon breaking TR symmetry. Red (blue) colour denotes spin-up (spin-down) electron states, the green circle denotes the spin-orbit mediated hybridization, and black lines indicate the emergent surface Fermi arcs.} 
	\label{fig:1}

\end{figure*}

Here we focus on a 2D topological ferromagnet that consists of two monolayers (MLs) Fe with in-plane magnetization, grown epitaxially on a W(110) substrate. It was recently predicted that bcc Fe is a topological metal as its electronic structure hosts emergent Weyl points and arc-like resonances on the (110) surface connecting them~\cite{gos15,gos18}. However, these non-trivial features usually do not affect macroscopic phenomena as they are located far away from the Fermi energy. By interfacing a thin Fe film with a heavy metal providing strong spin-orbit coupling (SOC), we promote this intrinsic ``hidden topology" and bring the associated topological features to the surface realm, with immediate consequences for macroscopic observables.

Owing to the different global properties of the topological metal Fe and the heavy-metal substrate W, the composite system is ideally suited to experimentally realize an interface-driven topological quantum phase in the presence of broken TR symmetry and strong SOC. We unveil the non-trivial topological state realized in this 2D magnet by studying the corresponding Fermi surface, using a recently available technique known as spin-resolved momentum microscopy~\cite{tus15,tus19}. Figure~\ref{fig:1}a presents the principle layout of the experimental geometry for our spin- and orbital-resolved photoemission study. The momentum microscope simultaneously collects photoelectrons over the full solid angle above the sample, such that spin- and momentum-resolved photoemission experiments over the whole Brillouin zone can be performed (see Methods)~\cite{tus15,tus19}. This technique gives comprehensive and intuitive access to the electronic structure of materials and
has advanced the frontiers in visualizing the orbital fingerprint throughout the whole Fermi surface by circular and linear dichroism~\cite{tus16,tus19}. Combined with the recent groundbreaking invention of an imaging spin detector~\cite{tus11,tus13}, this approach offers full access to detailed spin-resolved Fermi surfaces that were previously inaccessible by conventional photoelectron spectroscopy~\cite{tusche2018nonlocal,tus15}.

%
%
\begin{figure*}[t]
	\centering{\includegraphics[clip,width=15.8cm]{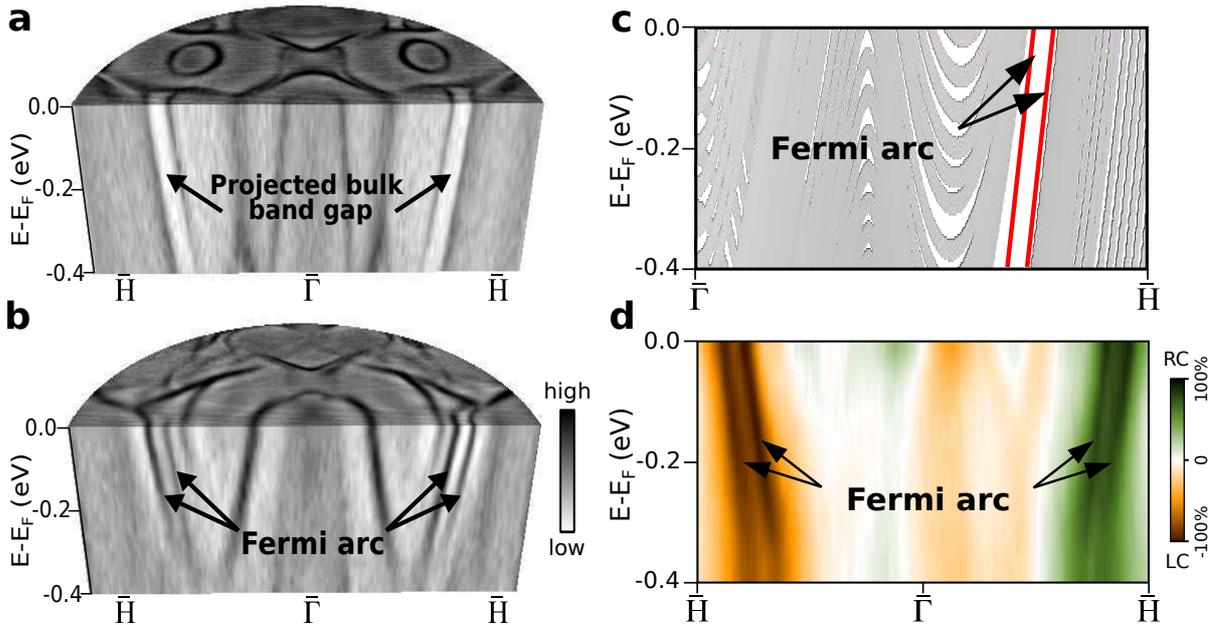}}
	\caption{\textbf{Fully helicity-selective surface arcs.}
		Band dispersions measured on (\textbf{a}) W(110) and (\textbf{b}) 2 MLs Fe/W(110) along $\overline{\mathrm{H}}-\overline{\Gamma}-\overline{\mathrm{H}}$ at the photon energy $h\nu$=50\,eV for the magnetization direction $M_{+y}$. \textbf{c}, Schematic diagram of the two co-propagating surface Fermi arcs (red lines) in the 2D topological ferromagnet, which appear inside the spin-orbit-induced  projected band gap (large white area) of the substrate. \textbf{d}, The measured circular dichroism reveals that the surface arcs in 2 MLs Fe/W(110) are 100\% helicity-selective, with opposite helicity between the two sides of the momentum map. The labels ``RC" and ``LC" denote right and left circular polarization of the light, respectively, and the colour code represents the magnitude of the circular dichroism signal.
		}

	\label{fig:2}
\end{figure*}

\begin{figure}[tb]
	\centering{\includegraphics[clip,width=7.0cm]{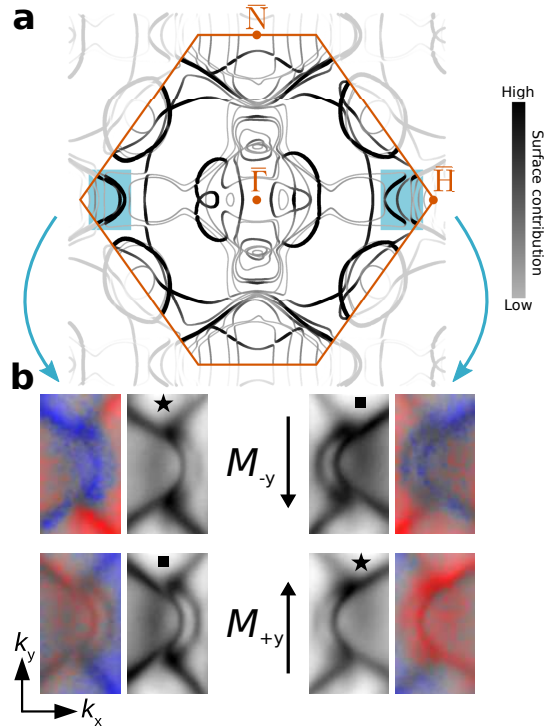}}
	\caption{\textbf{Fermi-arc switching induced by magnetization reversal.}
	     \textbf {a}, Theoretical Fermi surface of 2 MLs Fe/W(110) for the in-plane magnetization $M_{-y}$. Colours represent the degree of the states' localization in the magnetic layers, and the shaded regions highlight the emergent Fermi arcs.
		 \textbf{b}, Measured spin-resolved and spin-integrated surface arc states in the highlighted regions of (\textbf{a}) for the two magnetization directions $M_{+y}$ and $M_{-y}$ that are parallel and anti-parallel to the $y$ direction, respectively. The shape and the spin polarization of the strongly asymmetric Fermi arcs on opposite sides of the momentum map is interchanged if the magnetization direction is reversed as visualized by the black star and square symbols.} 
	\label{fig:3}
\end{figure}

Figure~\ref{fig:1}c shows the spin-resolved momentum map that we obtain at the Fermi energy, revealing the appearance of prominent surface states in the highlighted regions on both sides of the momentum map. These open arcs with a crescent-moon shape exhibit a high spin polarization. By comparing our experimental results to first-principles calculations (see Figs.~\ref{fig:3}a and Supplementary Fig.~S1), we identify these arcs as true surface states that exist as open segments only in a limited region of the surface Brillouin zone, and which terminate at the intersections with the interior states. Interestingly, the location of the surface Fermi arcs on 2 MLs Fe/W(110) corresponds to the region where two non-trivial Fermi surface electron pockets with opposite Chern number emerge in bulk bcc Fe (see Supplementary Fig.~S2)~\cite{gos15}. 
Thus, one can anticipate that the observed surface arcs are associated with these chiral band
degeneracies in Fe. The observation of emerging surface arc states in a 2D ferromagnet provides a rich
platform for exploring topological magnets beyond conventional Weyl semimetals.
As the observed open surface states share strong analogues with their cousins in 3D topological materials~\cite{Xu613,Hal14,min19,bra16,rao19,soluyanov2015type}, we refer to them as Fermi arcs in a 2D topological ferromagnet.

To understand the origin of these non-trivial states, we repeat the measurements for a reference system where the thin ferromagnet is replaced with a thick Fe film of twelve MLs. In that case, the surface Fermi arcs disappear and two states with opposite spin polarization emerge instead, crossing each other without any hybridization (see Fig.~\ref{fig:1}b). This bulk-like behaviour is drastically different from the electronic structure of the atomically thin magnetic films that are strongly affected by the sizable SOC of the substrate.

An important condition for the formation of the Fermi arc surface states is the topological transition
of the Fermi surface that is triggered by the strong SOC brought about by the heavy-metal
substrate. The changed topology of the Fe Fermi surface sheets can be understood as follows: due to the
presence of magnetic exchange interaction, the spin degeneracy is lifted in ferromagnetic Fe. In the limit of sufficiently large exchange splitting and negligible SOC,
majority and minority spin states intersect each other as shown in Fig.~\ref{fig:1}b. Thus, without SOC, the majority ($\uparrow$) and minority ($\downarrow$) bands are decoupled and can not interact.
	
As illustrated in Fig.~\ref{fig:1}d, the degeneracy at the intersection region is removed by
strong SOC, mediated by hybridization of the wave functions with the heavy-metal substrate tungsten.
As a result, the spin-orbit-induced local gap leads to a discontinuity of the formerly closed contours
of the majority and minority Fermi surface sheets of Fe, contravening that the iso-energy contour
at the Fermi level of any conventional metal needs to be closed by a well-defined quasiparticle
dispersion.

Unlike closed loop surface states of ordinary metals, where quasiparticles travel merely on the surface,
Haldane~\cite{Hal14} has shown that the Fermi arc surface states act as ``surface conduits'' that
adiabatically transfer quasiparticles between topologically-non-trivial sheets of the metallic Fermi
surface, which are otherwise disjoint. Thus, the open ends of the Fermi arcs are connected to these different
Fermi surface sheets
such that the incomplete surface band ends at the respective intersection points. This mechanism allows the system to maintain a common chemical potential across the apparently disconnected Fermi
surface sheets, and can provide an experimental diagnostic
evidence for the existence of a Weyl metal~\cite{Hal14,min19}. In 3D materials, this is further known to
result in highly non-trivial consequences on the spin- and orbital degrees of freedom in the
system~\cite{Hal14,min19}. As we demonstrate below, similar conclusions likewise apply in the
present case of a topological 2D ferromagnet.

As it is important to know whether the surface arcs lie in a band gap of the substrate, we compare the results of surface band structure measurements without and with the thin magnetic layer on top of the substrate. As shown in Fig.~\ref{fig:2}a, an energy gap is found in W(110) along the
$\overline{\mathrm{H}}-\overline{\Gamma}-\overline{\mathrm{H}}$ line, which
has been ascribed to a partial bulk band gap that is induced by strong SOC, thereby resembling the relativistic fundamental gap in topological insulators~\cite{Miyamoto2012,kut16,PhysRevB.95.245421,Elm17}. When an atomically thin Fe film is grown on the W(110)
substrate, the surface arcs with linear dispersion co-propagate
inside this projected bulk band gap (see Fig.~\ref{fig:2}b, c).
This result exhibits an analogy to the quantum anomalous Hall (QAH)
effect in doped topological insulators where chiral edge states co-propagate in the same
direction~\cite{Gan11}. In that case, an opposite handedness is observed at opposite side of Brillouin zone with
reversed velocity. Such a mechanism is general for realizing QAH phases with combined spontaneous
magnetic moments and SOC in the absence of an external magnetic
field~\cite{yu2010quantized,chang2013experimental}.

Figure~\ref{fig:3}a shows the calculated Fermi surface of 2ML Fe on W(110).
The colour in Fig.~\ref{fig:3}a represents the degree of states’ localization in the Fe layers. These dark coloured states are essentially derived from Fe, but can couple to the states of the substrate.
Thus, the wave functions of 2ML Fe hybridize and extend into the continuum of the Bloch states of the W substrate. 
This results in the Fermi surface contour being considerably different from the one of the corresponding freestanding Fe film, shown in Supplementary Fig.~S3a.
The calculation further reveals that the emergent Fermi arcs in the highlighted regions are located within the projected band gap of the bulk W(110) substrate.
The Fermi arcs therefore are characterized as true surface states, which have
no bulk correspondence, in contrast to the other ``interior" states in the 2ML
film.
As clearly evident, the substrate plays a crucial role in promoting the non-trivial features of the hidden topology
in Fe.
In particular, no Fermi arcs emerge for a freestanding 2ML Fe(110) film (see Supplementary Fig.~S3a). This result is in line with previous reports of
freestanding single monoloayer Fe films where no Fermi arcs were observed, likewise~\cite{nak06,buc11}.

%
%

%
%
\begin{figure*}[htb]
	\centering{\includegraphics[clip,width=16.3cm]{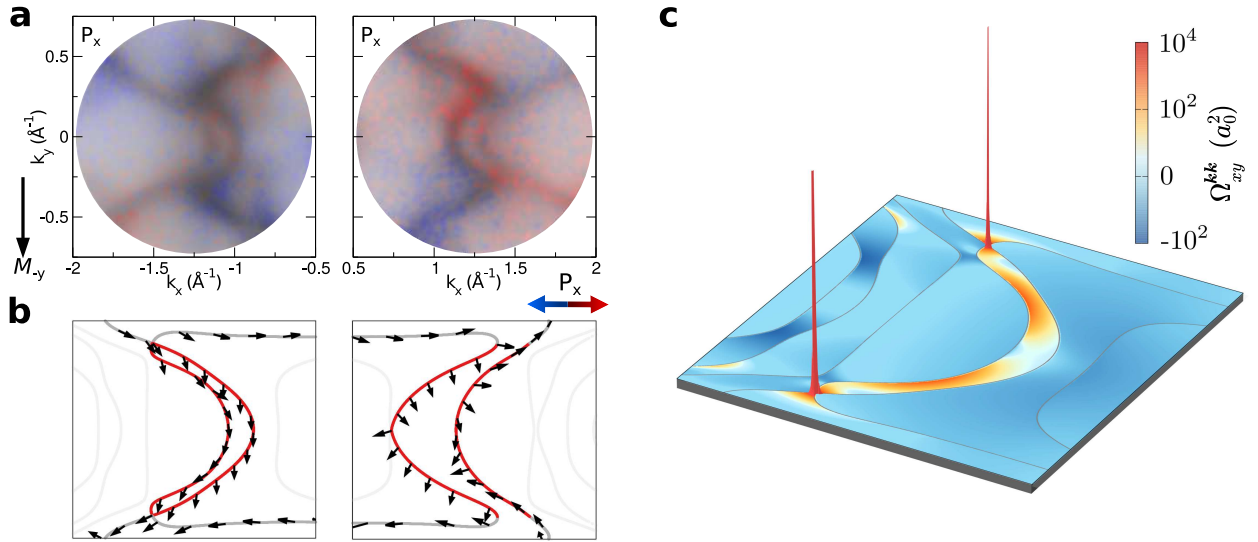}}
	\caption{\textbf{Spin texture and Berry curvature of the Fermi arcs in the 2D topological ferromagnet.}
		\textbf{a}, Measured spin-resolved Fermi arcs in 2 MLs Fe/W(110) on the left and on the right side of the momentum map. Colours indicate the in-plane spin component along the $x$ direction, which is orthogonal to the sample magnetization $M_{-y}$. \textbf{b}, Arrows denote the complete theoretical spin texture in momentum space, revealing a prominent non-collinearity for the Fermi arcs (red) as compared to the interior states (grey). \textbf{c}, Distribution of the theoretical momentum-space Berry curvature $\Omega_{xy}^{\textbf{kk}}$ of all occupied bands in 2 MLs Fe/W(110), around one of the pairs of Fermi arcs. A pseudo-logarithmic colour scale and a linear height field are used to represent the local geometrical curvature, and the Fermi surface is indicated by grey lines.
		}
	\label{fig:4}
\end{figure*}

Quantum phenomena that result from the breaking of TR symmetry offer an innovative platform for ``on-demand'' design and nonvolatile switching of physical properties by controlling the direction of the sample magnetization~\cite{Han17,jac19,Tok19}. To explore the magnetization response of the observed arc states, we performed momentum microscopy experiments for both magnetization orientations, parallel and anti-parallel
to the $y$ direction as denoted by M$_{+y}$ and M$_{-y}$, respectively. 
As apparent from Fig.~\ref{fig:3}, these orientations of the magnetization break the two-fold
symmetry of the (110) surface, leading to a Fermi surface that exhibits mirror symmetry with respect
to the $k_y=0$ mirror plane, whereas non-equivalent features occur with respect to the
$k_x=0$ plane. Remarkably, among all states at the Fermi surface, the open arcs show the most striking
asymmetry of the photoemission intensity and the shape
between the left and right side of the momentum map.
Such magnetization-dependent asymmetries require the presence of SOC\cite{Mly16}.
Here, the strong SOC is mediated through hybridization with substrate states in the regions of the Fermi arcs.
The pronounced $\mathbf k$-dependent relativistic splitting is then promoted by
both, the broken TR symmetry and missing spatial inversion symmetry at the interface. This mechanism
is reminiscent of the physical process that gives rise to the interfacial Dzyaloshinskii-Moriya interaction, leading to chiral and
non-collinear magnetic structures. By virtue of the spin-resolved data, Fig.~\ref{fig:3}b, a closer look at the surface arcs reveals that
the sign of their spin polarization $P_y$ changes upon magnetization reversal as the surface arc topology is interchanged between the two sides of the momentum map. 

In $3d$ metallic ferromagnets, most of the states are nearly pure spin states, namely, either spin-up (majority) or spin-down (minority) states with respect to the sample magnetization. 
Due to the broken inversion symmetry at interfaces with heavy-metal substrates that provide SOC, non-collinear spin textures may be favoured over collinear configurations. To determine the full spin texture of the surface arcs, we have measured in our momentum microscopy experiments also the in-plane spin polarization $P_x$ along the $x$ direction, which is orthogonal to the sample magnetization $M_{-y}$ of the 2D topological ferromagnet. According to our first-principles calculations, the states carry no spin polarization perpendicular to the film plane. As shown in Fig.~\ref{fig:4}a, the Fermi arcs exhibit a surprisingly sizable spin polarization $P_x$ that amounts to as much as $50\%$ of the signal $P_y$. 
A prominent variation of spin polarization is observed in the right pair of arcs, indicating a significant non-collinearity of the spin texture. By contrast, this non-collinearity is less pronounced for the left pair of arcs.
For comparison, we provide in Fig.~\ref{fig:4}b the theoretical spin texture of the corresponding surface arcs.
In good qualitative agreement with our experimental findings,
the theory confirms for the right arcs a pronounced deviation of the local spin orientation from the vertical magnetization direction.

In contrast to the Fermi arcs that appear in Weyl semimetals with TR symmetry~\cite{Xu613,min19}, our results for 2 MLs Fe/W(110) demonstrate that not only the shape, but also the \emph{spin texture} of the Fermi arcs is asymmetric with respect to the centre of the surface Brillouin zone due to the global breaking of TR symmetry. 
Clearly, we find that the full spin texture uncovers a prominent degree of non-collinearity in the Fermi arcs, distinguishing them from the regular ferromagnetic states with collinear spin polarization.
Since the evanescent wave functions of the surface
arcs inherit their velocity from the interior states of the film, they are attached tangentially as they approach the termination points, as observed in Fig.~\ref{fig:4}b.
Analogously, the complex interplay between topology and magnetism is well known to locally imprint also the orbital properties of individual states close to topological phase transitions~\cite{Niu19}. Indeed, our first-principles calculations in Supplementary Fig.~S4 demonstrate that the distinct topology of the observed arc-like surface states is not only correlated with the spin polarization but manifests as well in unique changes of the orbital character, as the corresponding states evolve into the termination points. Consequently, our experimental and theoretical results unveil
that the rapid variation in the spin and orbital texture reflects the abrupt transition from the
genuine surface arcs into the interior states at the intersection points, in close resemblance to
the orbital fingerprint of topological Fermi arcs in a Weyl semimetal~\cite{min19}.

\begin{figure*}[t]
	\centering{\includegraphics[clip,width=16.5cm]{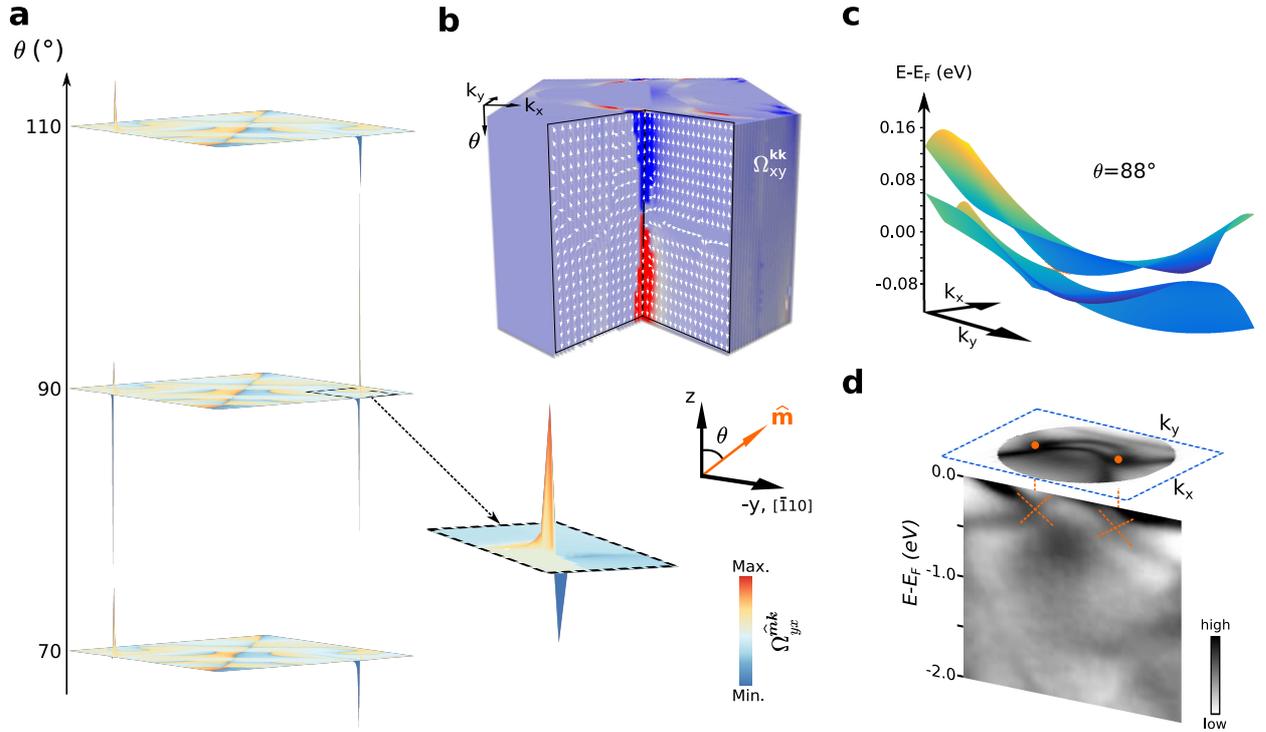}}
		\caption{\textbf{Mixed Weyl cones in a 2D topological ferromagnet.} 
			\textbf{a}, The mixed Berry curvature distribution $\Omega_{yx}^{\textbf{\^mk}}$ as a function of the magnetization direction ($\theta$); inset shows a magnified view near one end of the surface arc to highlight the sharp change in its sign across the projected mixed Weyl nodes.
			A logarithmic colour scale and a linear height field are used, where red (blue) denotes positive (negative) values. 
			\textbf{b}, Monopole-like field of the phase-space distribution of the Berry curvature
			$\Omega=(-\Omega_{yy}^{\textbf{\^{m}k}}, \Omega_{yx}^{\textbf{\^{m}k}}, \Omega_{xy}^{\textbf{kk}})$ near the end points of the Fermi arcs at $\theta=88^{\circ}$.
			Arrows indicate the direction of the Berry curvature field, and the colour scale from blue (negative) to red (positive) encodes the momentum Berry curvature $\Omega_{xy}^{\textbf{kk}}$
			in the composite space of $\mathbf{k}$ and $\theta$.
			\textbf{c}, Calculated band dispersion of the mixed Weyl cone emerging at $\theta=88^{\circ}$.
			\textbf{d}, Measured band dispersions along a straight cut through two intersection points indicated by orange dots.
			Surface Fermi arcs spanning in the $(k_x-k_y)$ plane connect the projection of the mixed Weyl nodes. The experimental observation of band crossings near the Fermi level agrees well with theoretical predictions in (\textbf{c}). The orange dashed lines are a guide to the eye.
		}
		\label{Fig5}
\end{figure*}

The Fermi-arc states in usual Weyl semimetals inherit  chiral properties from the topological nodal points that they connect. To explore whether the observed Fermi arcs in the present system with broken TR symmetry reflect this chiral nature as well, we carried out momentum-resolved photoemission experiments using left- and right-circularly polarized light, without changing the overall experimental geometry. Figure~\ref{fig:2}d presents our results for the circular dichroism (CD), which uncovers that the difference in excitation for light with different sense of circular polarization amounts to as much as 100\% for the arc states. As a consequence, photons with different helicity excite selectively the Fermi arcs either at the right side or at the left side of the momentum map.
The sign and magnitude of CD remains unchanged when the spin-texture is reversed.
This implies that the dichroic signal mainly originates from the orbital part of the wave functions,
being independent of the initial-state spin polarization. In contrast to the other states with relatively
broad CD spectral features due to coupling to continuum final states, the remarkably sharp and large
CD signal of the co-propagating surface arcs in the entire binding energy range can be primarily
attributed to the Fermi arc ground states.

In typical Weyl semimetals, the CD signal correlates with the chirality of nodal points, reflecting non-trivial global properties of the Bloch states in momentum space as encoded locally in the Berry curvature~\cite{ma2017direct,zhang2018optical}. The latter geometrical quantity is a key ingredient in the interpretation of the intrinsic contribution to the anomalous Hall effect. Transferring this concept to the considered case of a 2D topological ferromagnet, we point out that the regions with the largest CD signal in Fig.~\ref{fig:2}d, correspond exactly to the emergent Fermi arcs, and provide substantial contributions to the Berry curvature in momentum space, as we will see below.

In order to verify the microscopic insights revealed by our dichroism measurements, we performed theoretical first-principles calculations of the momentum-space Berry curvature for the considered 2D topological ferromagnet. Figure~\ref{fig:4}c presents the resulting distribution of the Berry curvature summed over all occupied bands, which underlies many macroscopic properties such as orbital magnetism and intrinsic Hall responses. While there are several minor background contributions, the most prominent features stem from states that form the observed Fermi arcs, with strong hot spots where the surface arcs connect to the attachment points, see also Supplementary Fig.~S5.

Conventional 3D Weyl materials are well described in the space of the 3D crystal momentum $(k_x,k_y,k_z)$,
where the Weyl nodes are located. In a 2D ferromagnet, as in the present case of 2ML Fe/W(110), this
description breaks down, since the momentum coordinate $k_z$ is not a good quantum number in the
low-dimensional system. Beyond this conventional Weyl picture, the emergence of topological points can be
understood when we consider the additional degree of freedom of the magnetization, that is already the
source of broken TR symmetry. In this description, the regular $(k_x,k_y,k_z)$ momentum space is replaced
by a mixed phase space $(k_x,k_y,\hat{\mathbf{m}})$, where $\hat{\mathbf{m}}$ denotes the magnetization
direction~\cite{Han17,Niu19}.
The observed pronounced change of Fermi surface states induced by changes of the magnetization, thus, can be interpreted as a direct signature of the topology of this \emph{mixed} phase space.
To investigate the origin of the open Fermi arcs at the surface of the 2D magnetic layer, we include the
magnetization direction and the mixed components of the Berry curvature into the topological analysis.

The magnetization direction $\textbf{\^{m}}=(\sin \theta, 0, \cos \theta)$ of a 2D magnet encloses the
angle $\theta$ with the $z$-axis.
As shown in Fig.~\ref{Fig5}a, the sign of the mixed Berry curvature ($\Omega_{yx}^{\textbf{\^{m}k}}$)
changes sharply across the intersection points with the Fermi arcs in the $k_x-k_y$ plane (see
Fig.~\ref{Fig5}a), while the momentum Berry curvature ($\Omega_{xy}^{\textbf{kk}}$) changes the
sign with the magnetization direction at $\theta\approx88^{\circ}$. The Berry curvature changes
sign when the topological phase transition occurs.
Figure~\ref{Fig5}b reveals that the Berry curvature field \pmb{$\Omega$} acquires a
monopole-like distribution around these mixed Weyl points, located at $\theta\approx88^{\circ}$.
This can be directly related to the presence of topological charges in the composite phase
space, which act as sources and sinks of the curvature field, and drive the emergence of the
Fermi arc states.

In order to identify the source of these topological charges we followed the evolution
of the electronic structure of the 2ML Fe film with the magnetization direction ($\theta$).
As shown in Fig.~\ref{Fig5}c, strongly tiled cones emerge at band touching points of electron
and hole pockets at $\theta\approx88^{\circ}$. These degeneracy points form the mixed Weyl
points that give rise the charges that we have observed in the mixed phase space in
Figs.~\ref{Fig5}a and \ref{Fig5}b.
As clearly visible in our calculations (Fig.~\ref{Fig5}c) and the experimental data (Fig.~\ref{Fig5}d),
the open-loop Fermi arcs are connected to the projections of these mixed Weyl nodes at
$\theta\approx88^{\circ}$ around the Fermi energy.

The presence of such emergent monopoles in the mixed phase space close to the Fermi energy has been
previously found as the source of a particularly strong Dzyaloshinskii-Moriya interaction
(DMI)~\cite{Han17,Niu19}. Our present results therefore provide an explanation for prior reports
of an exceptionally large DMI in 2 ML Fe/W(110)~\cite{Udv09,Zak10}, which thus can be directly related
to the presence of the non-trivial Fermi surface topology in this system.
Correspondingly, the exotic complex topology in the high-dimensional space boosts the variation of the spin and orbital character of Fermi arcs.  

As a consequence, these regions of large geometrical curvature constitute the most important sources of macroscopic effects related to charge transport and orbital properties, rendering the emergent Fermi arcs
the key ingredients in understanding these phenomena in 2D topological ferromagnets. Moreover, owing to
the distinct nature of the open arcs as compared to the trivial Fe and W(110) states, we expect that the former activate
additional current-induced field-like torques that act on the magnetization.

In summary, we have identified the appearance of open surface states in a 2D topological
ferromagnet, resulting from the complex interplay between an itinerant ferromagnet
and a heavy metal with strong spin-orbit coupling. Our results suggest that the occurrence of non-trivial Fermi
surface topologies is much more abundant in conventional magnetic material than
commonly expected. An advanced materials design that tailors the concurrence of
symmetry, exchange, and SOC can bring such hidden topologies to the surface, with
direct implications on exotic magneto-transport properties. Uncovering these principles
can promote innovative technologies based on topological spin-orbitronics, where
non-trivial topological states are controlled ``on demand" through the magnetization
direction.
\\



\section*{Methods}

\noindent\textbf{Spin-resolved momentum microscopy.}
Spin- and momentum-resolved photoelectron spectroscopy experiments were carried out at the NanoESCA beamline~\cite{wie11} of the Elettra synchrotron in Trieste (Italy), using p- and s-polarized photons at the photon energy $h\nu$=50\,eV. All measurements were performed while keeping the sample at a temperature of 130~K. Photoelectrons emitted into the complete solid angle above the sample surface were collected using a momentum microscope~\cite{tus15,tus19}. The momentum microscope directly forms an image of the distribution of photoelectrons as function of the lateral crystal momentum ($k_x$,$k_y$) that is recorded on an imaging detector~\cite{tus15,tus19}. 

An imaging spin filter based on the spin-dependent reflection of low-energy electrons at a W(100) single crystal~\cite{tus11} allows the simultaneous measurement of the spin polarization of photoelectrons in the whole surface BZ. Images were recorded after reflection at a scattering energy at 26.5\,eV and 30.5\,eV that changes the spin sensitivity $S$ of the detector between 42\% and 5\%, respectively. From these images, the spin polarization at every ($k_x,k_y$) point in a momentum map at certain binding energies
(e.g., Figs.~1 and~3-4 in the main manuscript) is derived along the procedure outlined in Refs.~\cite{tus11,tus13,tusche2018nonlocal}.

The W(100) crystal of the spin-filter was prepared by following the same procedure as described below for W(110). This standard procedure is known to lead to clean tungsten
surfaces~\cite{tus11,tus13}. At a pressure of $1\times 10^{-10}$\,mbar inside the
spin-filter chamber during the measurements, the analyser-crystal could be used for 2 hours. This time frame allowed for collecting 2D spin-resolved momentum maps at several energies~\cite{tusche2018nonlocal}.
\\

\noindent\textbf{Growth of the quasi-2D Fe monolayer on W(110).}
For the growth of 2 monolayers (MLs) Fe on W(110), the primary important step is to prepare a clean tungsten surface. The procedure for cleaning W(110) consists of two steps: (i) cycles of low temperature flash-heating (T$\sim$1200~K) at an oxygen partial pressure of $P_{O_{2}}=5\times10^{-8}$ mbar to remove the carbon from the surface, and (ii) a single high temperature flash (T$\sim$2400~K) to remove the oxide layer~\cite{tus13}. The cleanliness of the W(110) surface was checked by low energy electron diffraction (LEED) and Auger electron spectroscopy. Iron films with a thickness of 2 MLs were deposited in situ by molecular beam epitaxy from a high-purity Fe rod, heated by electron bombardment, onto the clean W(110) single crystal at a substrate temperature of 350~K. 

Before carrying out spin resolved momentum microscopy measurements, the samples were uniformly magnetized along the quantization axis of the spin filter, being the same direction as the easy magnetization direction ($\overline{\Gamma}-\overline{\mathrm{N}}$, $k_y$) of 2 MLs Fe films on W(110). We see that most electronic states are either of majority (spin up) or minority (spin down) character with the spin polarization along the $k_y$ axis as shown in Figs.~1 and 3 of the main manuscript. In order to measure the transverse component of the spin polarization parallel to the $k_x$ ($\overline{\Gamma}-\overline{\mathrm{H}}$) direction of the Fe films, we rotated the sample around 90$^{\circ}$. The $k_x$ direction of the sample then is parallel to the quantization axis of the spin filter (see Fig.~4 of the main manuscript).  
\\

\noindent\textbf{First-principles calculations.}
Using the full-potential linearized augmented-plane-wave (FLAPW) method as implemented in the \texttt{FLEUR} code~\cite{fleur}, we performed density functional theory calculations of 2 MLs Fe on 7 layers of W(110). We adopted the structural parameters given in Ref.~\cite{Heide2008}. Exchange and correlation effects were treated within the local-density approximation (LDA)~\cite{mjw}. The spin-orbit interaction was included self-consistently and the magnetization direction was chosen according to the experimentally observed easy axis $[1\bar 1 0]$. The plane-wave cutoff was chosen as $k_\mathrm{max}=4.2\,a_0^{-1}$ with $a_0$ as Bohr's radius, and the muffin-tin radii for Fe and W were set to $2.1\,a_0$ and $2.5\,a_0$, respectively.

To determine the Fermi surface with high precision, we calculated the eigenspectrum on a dense uniform mesh of $256$$\times$$256$ $\mathbf k$-points in the full Brillouin zone. The Fermi level crossings were then obtained by triangulation. To improve the agreement with experiment, we shifted the position of the Fermi level in our calculations downwards by $70$\,meV as compared to the theoretical value for the Fermi energy.

Based on the converged electronic structure, we used the wave-function information on a uniform mesh of $8$$\times$$8$ $\mathbf k$-points in order to construct a single set of $162$ maximally localized Wannier functions (MLWFs)~\cite{Marzari1997,Souza2001} out of $227$ Bloch states via the {\textsc{wannier}}{\footnotesize{90}} program~\cite{Mostofi2014,Freimuth2008}. In this procedure, we projected initially onto $d$-orbitals and $sp^3d^2$-hybrids, and we chose a frozen window that extends up to $1.5$\,eV above the Fermi level. The resulting MLWFs allow us to efficiently sample the full Brillouin zone via an accurate Wannier interpolation~\cite{Wang2006} that grants access to spin- and orbital-resolved properties, as well as to the Berry curvature.
\\




\


{

	\noindent\textbf{Acknowledgments} Y.-J.C.\ and C.T.\ thank the staff of Elettra for their help and hospitality during their visit in Trieste, and beamline staff M.~Jugovac, G.~Zamborlini, V.~Feyer (PGI-6, FZ-Jülich), and T.~O.~Mente\c{s} (Elettra) for their assistance during the experiment and providing the W(110) crystal. Y.-J.C., C.T., and C.S.\ gratefully acknowledge funding by the BMBF (Grant No. 05K19PGA). S.B.\ and Y.M.\ acknowledge support by the Deutsche Forschungsgemeinschaft (DFG) through the Collaborative Research Centre SFB 1238 (Project C1)  and Priority Programm SPP 2137. M.H., Y.M., and S.B.\ acknowledge funding from the DARPA TEE program through grant MIPR  (\#HR0011831554) from DOI. We also gratefully acknowledge the J\"ulich Supercomputing Centre and RWTH Aachen University for providing computational resources under projects jara0197, jias1f, and jiff40.\\
	
	
	


}

\clearpage

\end{document}


\title{\Large Spanning Fermi arcs in a two-dimensional magnet}
\author {Ying-Jiun Chen}
\email{yi.chen@fz-juelich.de}
\affiliation {Peter Gr\"unberg Institut, Forschungszentrum J\"ulich, 52425 J\"ulich, Germany}
\affiliation {Fakult\"at f\"ur Physik, Universit\"at Duisburg-Essen, 47057 Duisburg, Germany}

\author {Jan-Philipp Hanke}
\affiliation {Peter Gr\"unberg Institut, Forschungszentrum J\"ulich, 52425 J\"ulich, Germany}
\affiliation {Institute for Advanced Simulation, Forschungszentrum J\"ulich and JARA, 52425 J\"ulich, Germany}

\author {Markus Hoffmann}
\affiliation {Peter Gr\"unberg Institut, Forschungszentrum J\"ulich, 52425 J\"ulich, Germany}
\affiliation {Institute for Advanced Simulation, Forschungszentrum J\"ulich and JARA, 52425 J\"ulich, Germany}

\author {Gustav Bihlmayer}
\affiliation {Peter Gr\"unberg Institut, Forschungszentrum J\"ulich, 52425 J\"ulich, Germany}
\affiliation {Institute for Advanced Simulation, Forschungszentrum J\"ulich and JARA, 52425 J\"ulich, Germany}

\author {Yuriy Mokrousov}
\affiliation {Peter Gr\"unberg Institut, Forschungszentrum J\"ulich, 52425 J\"ulich, Germany}
\affiliation {Institute for Advanced Simulation, Forschungszentrum J\"ulich and JARA, 52425 J\"ulich, Germany}

\affiliation {Institute of Physics, Johannes Gutenberg University Mainz, 55099 Mainz, Germany}

\author {Stefan Bl\"ugel}
\affiliation {Peter Gr\"unberg Institut, Forschungszentrum J\"ulich, 52425 J\"ulich, Germany}
\affiliation {Institute for Advanced Simulation, Forschungszentrum J\"ulich and JARA, 52425 J\"ulich, Germany}

\author {Claus M. Schneider}
\affiliation {Peter Gr\"unberg Institut, Forschungszentrum J\"ulich, 52425 J\"ulich, Germany}
\affiliation {Fakult\"at f\"ur Physik, Universit\"at Duisburg-Essen, 47057 Duisburg, Germany}
\affiliation {Department of Physics, University of California Davis, Davis, CA 95616, USA}

\author {Christian Tusche}
\email{c.tusche@fz-juelich.de}
\affiliation {Peter Gr\"unberg Institut, Forschungszentrum J\"ulich, 52425 J\"ulich, Germany}
\affiliation {Fakult\"at f\"ur Physik, Universit\"at Duisburg-Essen, 47057 Duisburg, Germany}


\maketitle

\subsection{Photon-energy independent surface Fermi arcs}
\begin{figure*}[h]
	\center{\includegraphics[clip,width=11.0cm]{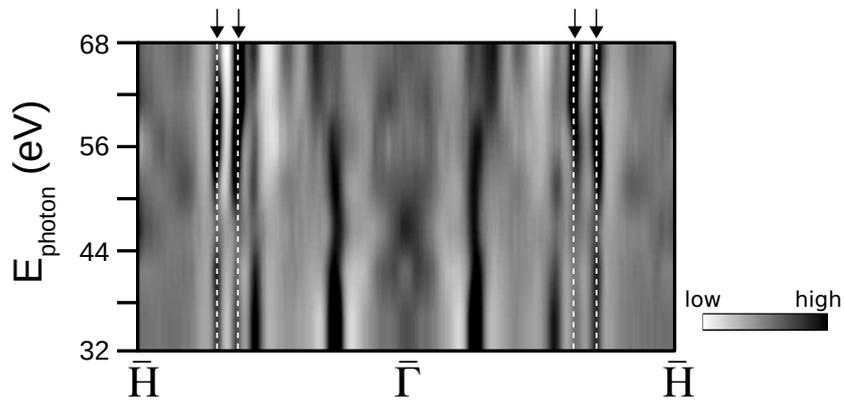}} 
	\caption{\textbf{Photon-energy ($k_z$) independent surface arc states.}
		Photoemission spectral intensity map at $E_F$ along $k_\parallel, y=0$ ($\overline{\mathrm{H}}-\overline{\Gamma}-\overline{\mathrm{H}}$)
		for $s$-polarized light with a photon energy range of 32-68\,eV that probes the various $k_z$ wave vectors perpendicular to the surface. The observed negligible $k_z$ dispersion of Fermi arcs denoted by the arrows confirms their 2D nature, showing a clear signature of surface states. White dashed lines serve as a guide to the eye.}
	\label{FigS1}
\end{figure*}

In order to characterize the dimensionality of the observed Fermi arcs, we carried out momentum microscopy experiments at different photon energies that allows us to probe the various $k_z$ wave vectors perpendicular to the surface. Figure~\ref{FigS1} shows the photoemission spectral intensity at $E_F$ along $k_\parallel, y=0$ ($\overline{\mathrm{H}}-\overline{\Gamma}-\overline{\mathrm{H}}$) with different photon energies. The lack of the $k_z$ dispersion confirms the 2D nature of the Fermi arcs (denoted by arrows), demonstrating a clear character of surface states. In contrast, highly dispersive bands around $k_\parallel$=0 are evident. This observation agrees well with our first-principles calculations in Fig.~3.

\subsection{Measured Fermi surface of ``bulk'' \lowercase{bcc} F\lowercase{e} surrounding topological pockets}
\begin{figure*}[h]
	\center{\includegraphics[clip,width=17.5cm]{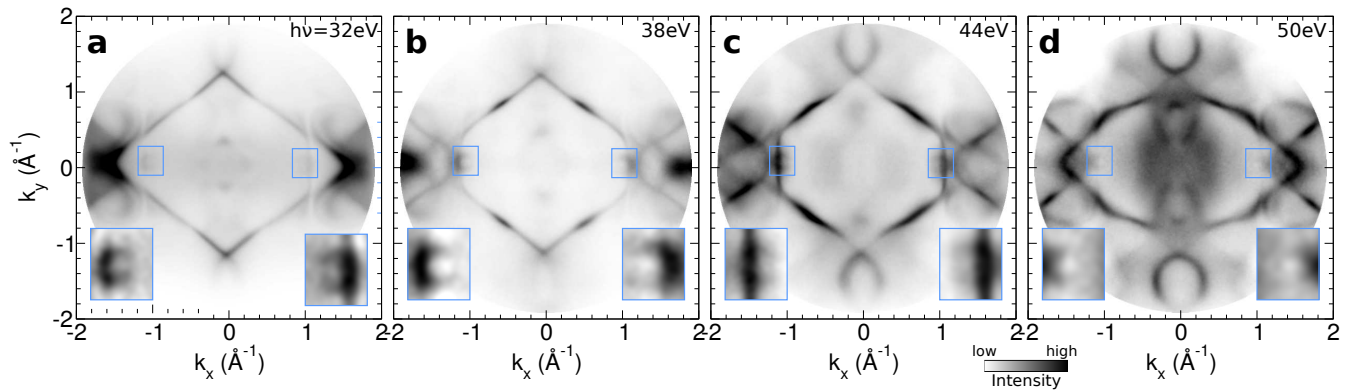}}
	\caption{\textbf{Topological electron pockets in ``bulk'' bcc Fe.} Measured 2D photoemission momentum map at the Fermi energy $\textit{E}_{F}$ for a 12 MLs Fe(110) film measured using s-polarized light with a photon energy of \textbf{(a)} 32\,eV, \textbf{(b)} 38\,eV, \textbf{(c)} 44\,eV and \textbf{(d)} 50\,eV, surrounding the two topological electronic pockets denoted by the blue rectangles. The inset shows the electron pockets at higher magnification. The intensity is displayed on a linear grey scale.
	}
	\label{FigS2}
\end{figure*}

It was recently predicted that bcc Fe
is  a  topological  metal with two disconnected Fermi surface sheets carrying opposite Chern numbers~\cite{gos15,gos18}. In order to identify the two non-trivial electron pockets in the Brillouin zone (BZ), measurements for a thick Fe(110) film of twelve MLs were performed. As highlighted in Fig.~\ref{FigS2}, the location and topology of the non-trivial electron pockets are identified at the Fe(110) surface, in a region corresponding to the location of the emergent surface Fermi arcs in 2 MLs Fe/W(110). Note that one would not be able to identify these two non-trivial electron pockets at the Fe(100) surface because they would be overlapping with projected bulk bands~\cite{gos18}.

\subsection{Orbital texture}
We present in Fig.~\ref{FigS4} the results of our first-principles calculations for the orbital texture around the Fermi arcs that emerge in 2 MLs Fe/W(110) with in-plane magnetization. The distinct topology of these features correlates with the changes of magnitude and direction in the in-plane orbital angular momentum $\mathbf L_\parallel$ of the surface arc states as they evolve into the intersection points. This finding is consistent with a recent theoretical~\cite{Niu19} and experimental~\cite{min19} study uncovering that the non-trivial interplay between magnetism and topology imprints locally also on the orbital properties of the electronic states in momentum space.

\subsection{Berry curvature}
The Berry curvature in momentum space, Eq.~\eqref{eq:berry}, is the driving force behind many macroscopic phenomena. Therefore, as shown in Fig.~\ref{FigS5}, we evaluate from first principles the distribution of the Berry curvature $\Omega_{xy}^{\textbf{kk}}$ around the Fermi arcs on the two opposite sides of the momentum map. Irrespective of a finite overall background, the most important contributions stem from points where the arc surface states evolve into the intersection points (note the pseudo-logarithmic scale in Fig.~\ref{FigS5}). The strong asymmetry in the shape of the Fermi arcs on opposite sides is also present in the distribution of $\Omega_{xy}^{\textbf{kk}}$.

\begin{equation}
\Omega_{xy}^{\textbf{kk}}(\mathbf k) = -2\,\mathrm{Im}\sum_n^\mathrm{occ}\sum_{m\neq n} \frac{\Big\langle u_{n\mathbf k} \Big| \frac{\partial H_{\mathbf k}}{\partial k_x} \Big| u_{m\mathbf k} \Big\rangle \Big\langle u_{m\mathbf k} \Big| \frac{\partial H_{\mathbf k}}{\partial k_y} \Big| u_{n\mathbf k} \Big\rangle}{(E_{n\mathbf k}-E_{m\mathbf k})^2} \, .
\label{eq:berry}
\end{equation}
Here, $u_{n\mathbf k}$ is an eigenstate of the lattice-periodic Hamiltonian $H_{\mathbf k}$ with the energy $E_{n\mathbf k}$, and the sum over $n$ is performed over all occupied bands with crystal momentum $\mathbf k$.	

The formation of the open Fermi arcs in two-dimensional ferromagnets is ascribed to the non-trivial geometry of the mixed phase space of $\textbf{k}$ and $ \textbf{\^{m}} $ in terms of the so-called mixed Berry curvature $\Omega_{ij}^{\textbf{\^{m}k}}$, Eq.~\eqref{eq:mixberry}, of all occupied states:

\begin{equation}
	\Omega_{ij}^{\textbf{\^{m}k}}(\mathbf k) = -2{\mathbf{\hat{e}_i}}\cdot\,\mathrm{Im}\sum_n^\mathrm{occ}\Big[\textbf{\^{m}}\times\Big\langle  \frac{\partial u_{\mathbf k n}}{\partial \textbf{\^{m}}} \Big| \frac{\partial u_{\mathbf k n}}{\partial k_j}\Big\rangle\Big] \, .
	\label{eq:mixberry}
	\end{equation}
	which incorporate derivatives of lattice-period wave functions $u_{\mathbf k n}$ with respect to both crystal momentum $\mathbf k$ and magnetization $\textbf{\^{m}}$. Here $\mathbf{\hat{e}_i}$ denotes the $i$th Cartesian unit vector.
	We used the Wannier interpolation that we generalized to treat crystal momentum and magnetization direction on an equal footing in order to evaluate the Berry curvature $\Omega^{\textbf{kk}}$ and $\Omega^{\textbf{\^{m}k}}$.

\subsection{Mixed topology in 2 ML Fe/W(110)}
The combination of a complex geometry in real and momentum spaces manifests the non-trivial mixed topology in low-dimensional magnets~\cite{Han17,Niu19}. The concept of a mixed topology was introduced by formally replacing one of the momentum variable with the magnetization direction, specified by an angle $\theta$ in the usual Weyl Hamiltonian. This results in the low-energy description of the system in the composite phase space of entangled crystal momentum $\textbf{k}=(k_x,k_y)$ and magnetization direction $\theta$ by $H=\nu_xk_x\sigma_x+\nu_yk_y\sigma_y+\nu_\theta\theta\sigma_z$, where $\theta$ is the angle that the magnetization makes with the $z$-axis~\cite{Han17,Niu19}.

As demonstrated in Fig.~5, 
to characterize the topological properties in the composite phase space spanned by $k_x$, $k_y$, and $\theta$, the momentum Berry curvature $\Omega_{xy}^{\textbf{kk}}$ and the mixed Berry curvature $\Omega_{ij}^{\textbf{\^{m}k}}$ are calculated around the Fermi arcs with respect to the magnetization direction, $\theta$. As shown in (a), we find the strong dependence of the Berry curvature on the magnetization direction and in-plane momentum.
Remarkably, both Berry curvature $\Omega_{ij}^{\textbf{\^{m}k}}$ and $\Omega_{xy}^{\textbf{kk}}$  show the maximum magnitude for $\theta\approx90^{\circ}$.
Essentially, the sign of the mixed Berry curvature ($\Omega_{yx}^{\textbf{\^{m}k}}$) changes sharply across the intersection points in the $k_x-k_y$ plane where the surface arcs attach to the projected Weyl nodes.
As a result, the Berry curvature field \pmb{$\Omega$} acquires a monopole-like distribution around the
mixed Weyl points. Figure~5b shows the corresponding vector field, where arrows indicate the direction
of the curvature field and the colour scale encodes the momentum Berry curvature
$\Omega_{xy}^{\textbf{kk}}$. We find sources of the curvature field at a magnetization
angle $\theta\approx88^{\circ}$. This characteristic field distribution is related to the
presence of topological charges at the mixed Weyl points \cite{Han17}.

Figure~5c shows the band dispersions of valence- and conduction-band states close to the Berry curvature peaks. By applying a detailed evaluation with respect to the magnetization direction $\theta$, we find that two bands only touch at $\theta\approx88^{\circ}$, near the Fermi level. This angle is remarkably close to the magnetization direction in the experiment ($\theta=90^{\circ}$), pointing approximately along the $-y$ direction.
Figure~5d shows the measured spectral function, where the degeneracy points appear at a slightly lower binding energy of 200\,meV below the Fermi level. The experimental data further confirms that the observed 
degeneracy points coincide with the open end of the Fermi arcs in the $k_x-k_y$ momentum space plane of the surface Brillouin zone.

The observation of a monopole-like distribution of the Berry curvature field constitute a strong signature of the topological phase transition and the presence of topological charges around the
end  points of the open surface arcs in a mixed phase space $(k_x,k_y,\hat{\mathbf{m}})$.
The non-trivial band structure of ultrathin 2D hybrid magnets proves to be a novel arena for exploring new topology physics beyond conventional 3D Weyl semimetals.
\\

\begin{figure*}[h]
	\center{\includegraphics[clip,width=16.0cm]{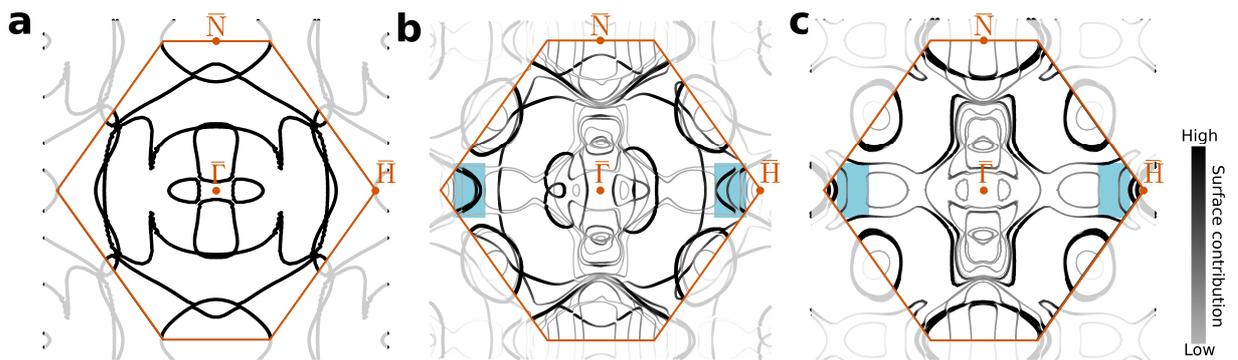}}
	\caption{\textbf{Theoretical Fermi surface of free-standing Fe(110) monolayers.} 
		Theoretical Fermi surface of (\textbf{a}) a free-standing 2ML Fe(110), (\textbf{b}) 2ML Fe/W(110), and (\textbf{c}) bulk W(110). The shaded blue regions highlight the emergent Fermi arc surface states in 2 ML Fe/W(110), which is located within the band gap of bulk W(110) and is absent in a free-standing film.
	}
	\label{FigS3}
\end{figure*}

\begin{figure*}[h]
	\center{\includegraphics[clip,width=16.0cm]{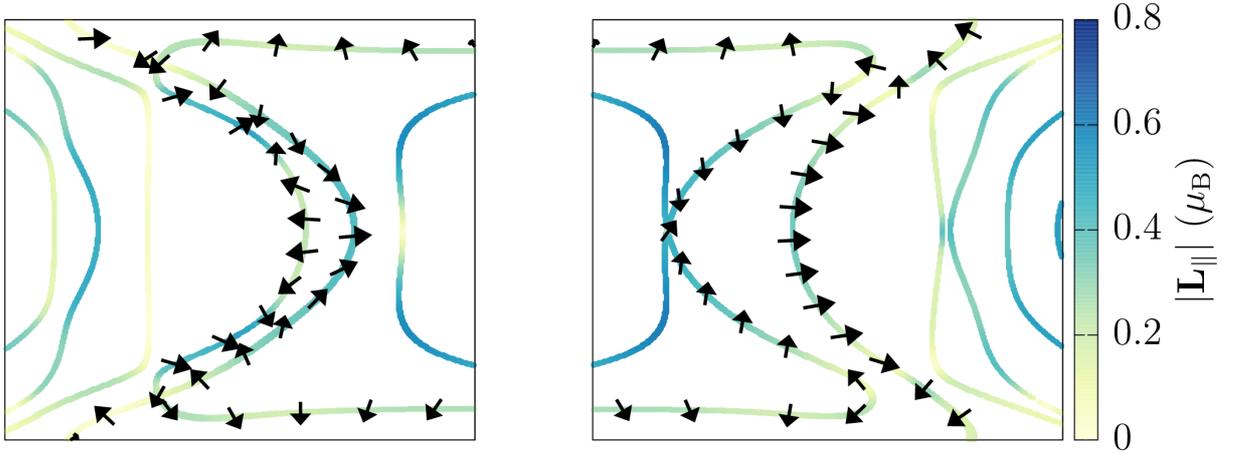}}
	\caption{\textbf{Orbital texture around the Fermi arcs in 2 MLs Fe/W(110).} The colour scale represents the magnitude of the local in-plane orbital moment $\mathbf L_\parallel$ in units of Bohr magneton, and unit vectors indicate its direction in the regions of the Fermi arcs on opposite sides of the momentum map.
	}
	\label{FigS4}
\end{figure*}

\begin{figure*}[t]
	\center{\includegraphics[clip,width=16.0cm]{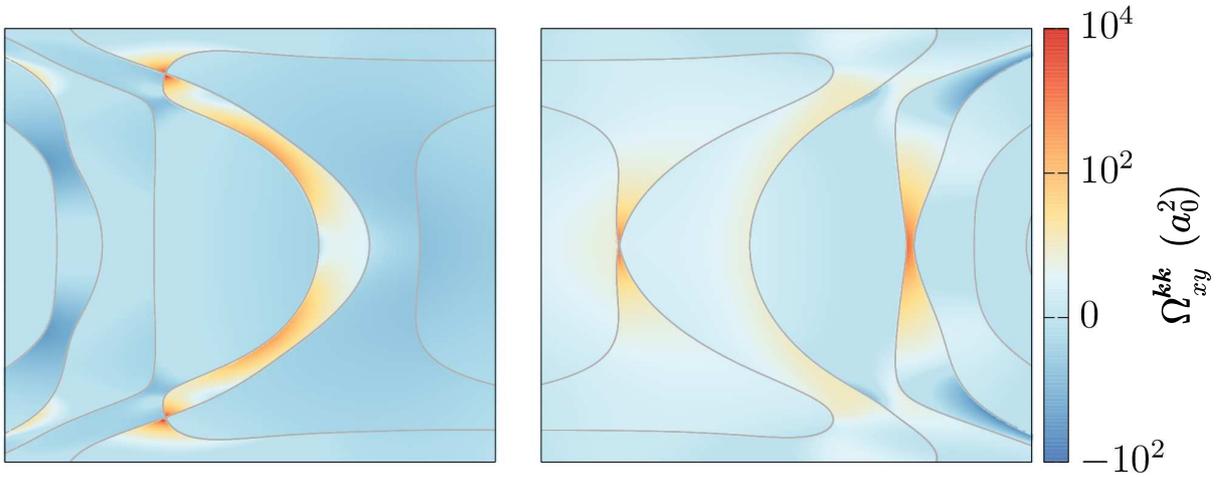}}
	\caption{\textbf{Berry curvature around the Fermi arcs in 2 MLs Fe/W(110).} 
		\textbf{a}, The local geometrical curvature $\Omega_{xy}^{\textbf{kk}}$ in the regions of the Fermi arcs is indicated by a pseudo-logarithmic colour scale. Grey lines represent the Fermi surface contour on the two sides of the momentum map.
	}
	\label{FigS5}
\end{figure*}

\clearpage

